\documentclass[%
reprint,
amsmath,amssymb,
 ]{revtex4-1}

\usepackage{graphicx}
\setcitestyle{super}

\begin{document}

\title{\Large\bfseries\noindent\sloppy \textsf{1 GHz dual-comb spectrometer for fast and broadband measurements}}

\author{Thibault~Voumard$^{1}$}
\author{John~Darvill$^{1}$}
\author{Thibault~Wildi$^{1}$}
\author{Markus~Ludwig$^{1}$}
\author{Christian~Mohr$^{1}$}
\author{Ingmar~Hartl$^{1}$}
\author{Tobias~Herr$^{1,2}$}
\email{tobias.herr@cfel.de}
\affiliation{
$^1$Deutsches Elektronen-Synchrotron DESY, Notkestr. 85, 22607 Hamburg, Germany \\
 $^2$Physics Department, Universität Hamburg UHH, Luruper Chaussee 149, 22761 Hamburg, Germany
}

\maketitle
\textbf{Dual-frequency comb spectroscopy permits broadband precision spectroscopic measurements with short acquisition time. A dramatic improvement of the maximal spectral bandwidth and the minimal measurement time can be expected when the lasers’ pulse repetition rate is increased, owing to a quadratic dependence (Nyquist criterion). Here, we demonstrate a dual-comb system operating at a high repetition rate of 1~GHz based on mature, digitally-controlled, low-noise mode-locked lasers. Compared to conventional lower repetition rate ($\sim$100~MHz) oscillators, this represents a 100-fold improvement in terms of the Nyquist criterion, while still providing adequate spectral sampling even for trace gas absorption fingerprints. Two spectroscopy experiments are performed with acquisition parameters not attainable in a 100~MHz system: detection of water vapor absorption around 1375~nm, demonstrating the potential for fast and ambiguity-free broadband operation, and real-time acquisition of narrow gas absorption features across a spectral span of 0.6~THz (600 comb lines) in only 5~$\mu$s. Additionally, we show high mutual coherence of the lasers below the Hz-level, generating opportunities for broadband spectroscopy even with low-bandwidth detectors such as mid-infrared, imaging or photo-acoustic detectors.}


\section{Introduction}
Dual-frequency comb spectroscopy (DCS) uses detection of multiheterodyne interferograms between two frequency combs with slightly different pulse repetition rates for coherent mapping of an optical spectrum into the radio frequency (RF) domain \cite{coddington_dual-comb_2016, picque_frequency_2019, schiller_spectrometry_2002, keilmann_time-domain_2004, coddington_coherent_2008, coddington_coherent_2010, link_dual-comb_2017, millot_frequency-agile_2016, newbury_sensitivity_2010, schliesser_mid-infrared_2012}.  This method can offer both broadband spectral coverage and rapid acquisition, as described and ultimately limited by the Nyquist criterion 
\begin{equation}
    \Delta \nu \leq \frac{f_\mathrm{r}(f_\mathrm{r} + \Delta f_\mathrm{r})}{2\Delta f_\mathrm{r}},
    \label{eq:Nyquist_1}
\end{equation}
where $\Delta \nu$ is the maximal optical bandwidth that can be mapped from the optical to the RF domain without ambiguity, $f_\mathrm{r}$ is the repetition rate of one of the two comb lasers and $\Delta f_\mathrm{r}$ is the repetition rate difference between the two lasers.

\begin{figure}[t]
    \centering
    \includegraphics[width=0.45\textwidth]{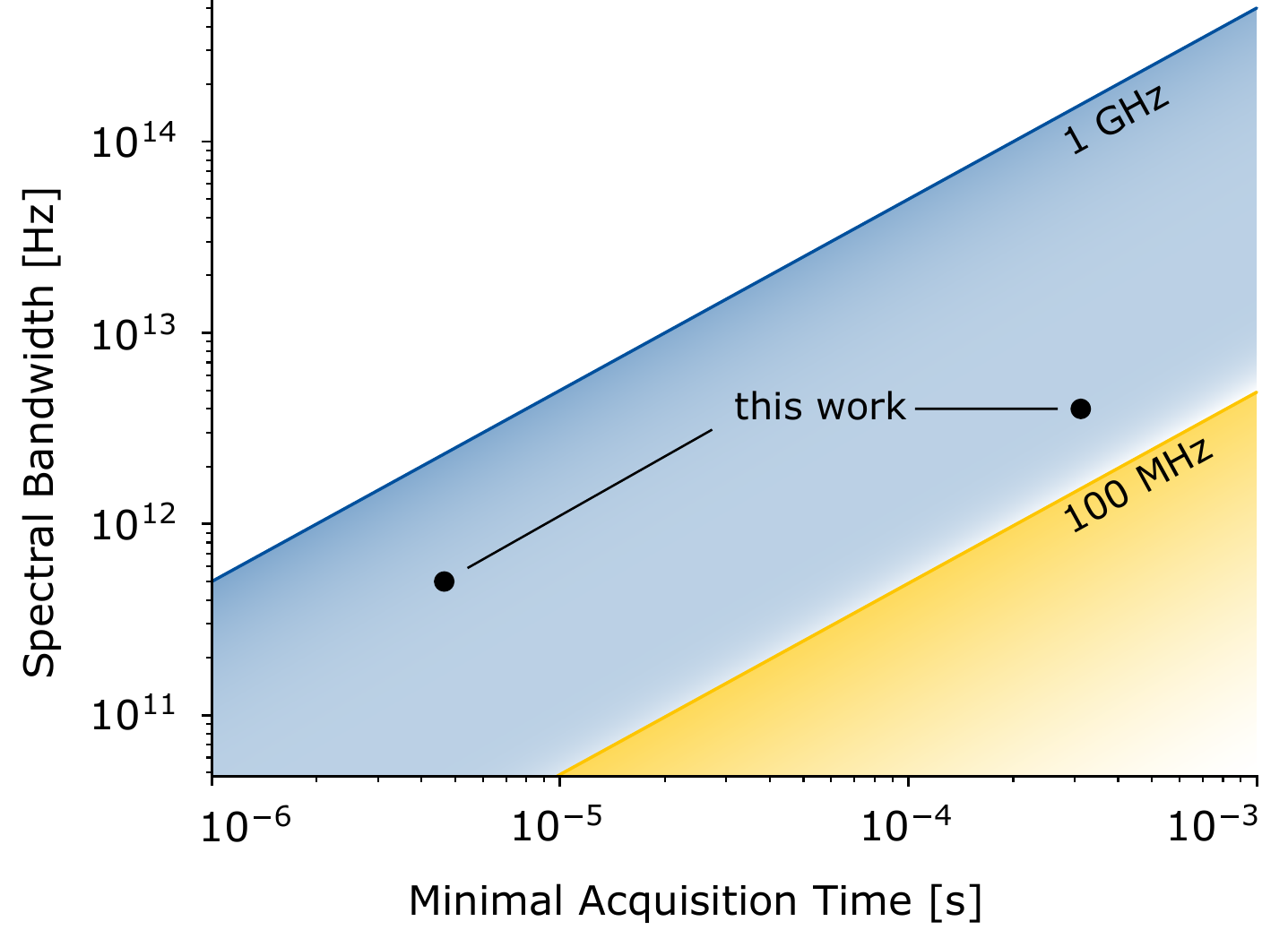}
    \caption{\textbf{Nyquist criterion.} Illustration of the Nyquist criterion for dual-comb spectroscopy in terms of the minimal measurement time and the maximal spectral bandwidth. The yellow area is accessible with a 100~MHz system, whereas a 1~GHz system can, in addition to the yellow area, operate in the blue-shaded parameter space. The parameters of the two experiments demonstrated in this work are indicated by the black dots (outside of the Nyquist range of a 100~MHz systems).}
    \label{nyquist}
\end{figure}

Broadband dual-comb spectroscopy \cite{okubo_ultra-broadband_2015, ycas_high-coherence_2018, zolot_direct-comb_2012} can thus be achieved in two ways: reducing $\Delta f_\mathrm{r}$, or increasing $f_\mathrm{r}$. The first approach has been pursued in dual-comb systems with high-mutual coherence \cite{gu_passive_2020, chen_phase-stable_2018, nakjima_high-coherence_2019, hebert_highly_2018, kayes_free-running_2018, liao_dual-comb_2018} with $\Delta f_\mathrm{r}$ as low as a few Hertz. Further reduction of $\Delta f_\mathrm{r}$ would require even higher-mutual coherence between the dual-comb lasers, which is challenging due to the inherent noise of laser oscillators \cite{haus_noise_1993}.
The second approach, i.e. increasing the repetition rate $f_\mathrm{r}$, is attractive due to the favorable quadratic scaling of $\Delta \nu \propto f_\mathrm{r}^2$.

Moreover, as the minimal acquisition time $\tau_\mathrm{min}$ for one optical spectrum is given by $\tau_\mathrm{min} = \Delta f_\mathrm{r}^{-1}$ (corresponding to the recording of a single interferogram), it follows from Eq.~\ref{eq:Nyquist_1} that $\tau_\mathrm{min} \propto f_\mathrm{r}^{-2}$. Hence, for a fixed optical bandwidth $\Delta\nu$, a higher repetition rate also reduces the minimal acquisition time quadratically. High acquisition rates are of interest as they enable spectroscopy of fast transient processes e.g. in high-throughput screening, flow-cytometry or plasma physics \cite{wang_fast_2019, kameyama_fast_2021, luo_fast_2019, abbas_time-resolved_2019}.

\begin{figure*}
    \centering
    \includegraphics[width=0.95\textwidth]{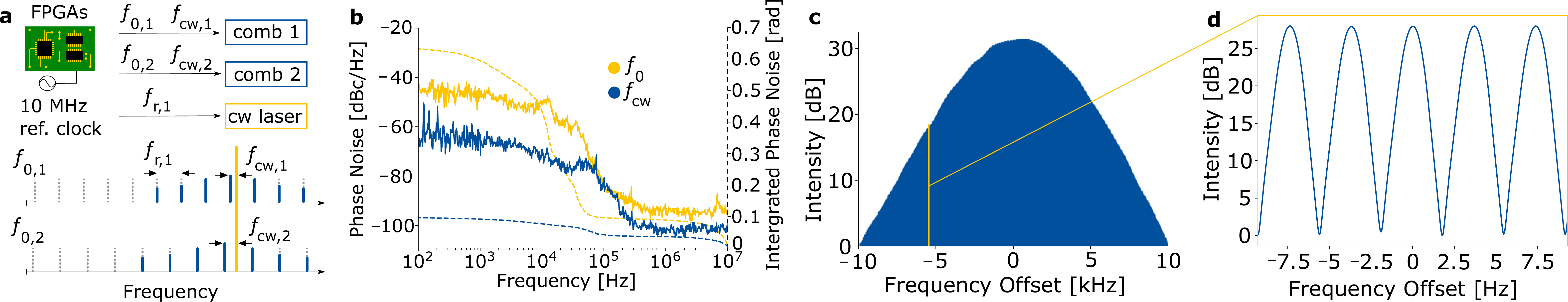}
    \caption{\textbf{Experimental setup and mutual coherence. a} Schematic of the field programmable gate array (FPGA)-based digital locking scheme for a high-mutual coherence, fully self-referenced dual-comb spectrometer. One line of each comb is phase locked to a continuous wave (CW) laser ($f_{\mathrm{CW},1,2}$). The CW laser's frequency is controlled to keep the repetition rate of comb~1 $f_{\mathrm{r},1}$stable against a radio frequency reference (10~MHz clock). Both carrier envelope offset frequencies $f_{0,i}$ are directly stabilized against the frequency reference.  \textbf{b} Residual phase noise power spectral density (solid curves) and phase noise integrated from 10~MHz (dashed curves) for phase lock between a comb and the CW laser (blue, offset frequency $f_\mathrm{CW,1}=250$~MHz) and carrier envelope offset phase lock (yellow, offset frequency $f_\mathrm{0,1}=390$~MHz). \textbf{c} Radio-frequency multi-heterodyne spectrum obtained by mixing the stabilized combs on a photodetector. The zoom-in \textbf{d} shows individual multiheterodyne beatnotes spaced by $\Delta f_\mathrm{r}\sim 4$~Hz with a 1~Hz resolution-bandwidth-limited linewidth, demonstrating the sub-Hz mutual coherence of the dual-comb system.}
    \label{lock_scheme}
\end{figure*}

Further, the signal-to-noise ratio (SNR) also scales with $f_\mathrm{r}$ for a given spectral bandwidth as the optical power is distributed among fewer comb lines \cite{newbury_sensitivity_2010}. The trade-off resulting from a higher repetition rate is a reduction in the number of spectral sampling points resolution. However, a 1 GHz repetition rate (i.e. a comb line spacing of 1~GHz) is sufficient to adequately sample even narrow gas absorption features at atmospheric pressure. Alternatively, the number of spectral sampling points can be increased by scanning the combs' repetition rates or offset frequencies at the cost of longer measurement time \cite{gianella_high-resolution_2020, ren_mid-infrared_2020}. 

High-repetition rate dual comb systems are more challenging to implement than their lower repetition rate counterparts due to lower pulse peak power. Previous demonstrations of high-repetition rate combs often relied on electro-optic combs \cite{carlson_broadband_2020, voumard_ai-enabled_2020, wildi_photo-acoustic_2020} where GHz-repetition rates are readily available but spectral broadening is difficult. In contrast, microresonator-based combs can achieve extremely high repetition rate and broadband spectral coverage but only provide very coarse spectral sampling \cite{suh_microresonator_2016}. Previous work with mode-locked lasers includes narrowband spectroscopy with Ytterbium-doped 1~GHz fiber-laser \cite{hartl_rapidly_2009} and 1~GHz Titanium-Sapphire lasers for dual-comb Raman spectroscopy \cite{mohler_dual-comb_2017}. Recently, a 1~GHz dual-comb system based on cavity-filtered 100~MHz erbium-doped fiber lasers has been implemented \cite{hoghooghi_11-s_2021}, acquiring a 750~GHz wide spectral span in only 11~$\mu$s highlighting the potential of high-repetition rate mode-locked laser systems. Although not yet reaching large spectral span, entirely chip-integrated 1~GHz dual-combs have been demonstrated highlighting the potential for low-cost and widespread use of such systems \cite{gasse_dual-comb_2021}.

Here, we demonstrate a dual-comb spectrometer based on two commercially available solid-state 1~GHz mode-locked lasers (1.56~$\mu$m central wavelength), offering a 100-fold improvement of the Nyquist criterion compared to established 100~MHz-based systems. We show a broadband spectral window without ambiguity in the optical-to-RF frequency mapping as well as fast spectroscopy with $\mu$s-scale acquisition time, both in regimes not attainable in a 100~MHz-based system (Fig.~\ref{nyquist}).  Moreover, high-mutual coherence (sub-Hz multiheterodyne beatnote width) is demonstrated providing optical-to-RF compression factors of $> 10^9$. All feedback loops are implemented via a compact and flexible FPGA-based digital locking scheme \cite{preuschoff_digital_2020, sinclair_invited_2015, neuhaus_pyrpl_2017}.

\section{Results}
\textbf{Setup.} The dual-comb spectrometer builds upon two commercially available low-noise solid-state 1~GHz mode-locked lasers with tunable repetition rate and offset frequency, with a center wavelength of 1560~nm. The combs are amplified to $\sim$900~mW of average optical power and broadened to an octave-spanning supercontinuum in a highly-nonlinear fiber (HNLF) supporting the formation of a Raman soliton. The combs' carrier-envelope offset frequencies are detected through f-2f interferometers (similar to ref.~\cite{lesko_fully_2020}) and stabilized to a 10~MHz RF standard by modulation of the pump currents. High-mutual coherence between both comb sources is achieved by phase-locking one line of each comb to a 1560~nm continuous-wave (CW) fiber-laser ($\sim$10~kHz linewidth) via piezo-actuators. In turn, the CW laser is (slowly) actuated such that the repetition rate of one comb $f_\mathrm{r,1}$ is stabilized and linked to the same 10~MHz RF reference. This implies that the CW laser's frequency as well as the repetition rate of the second comb are also stabilized and linked to the RF reference. In total, these five phase locks yield fully self-referenced dual-frequency combs with high-mutual coherence (Fig.~\ref{lock_scheme}a). 

\begin{figure*}
    \centering
    \includegraphics[width=0.95\textwidth]{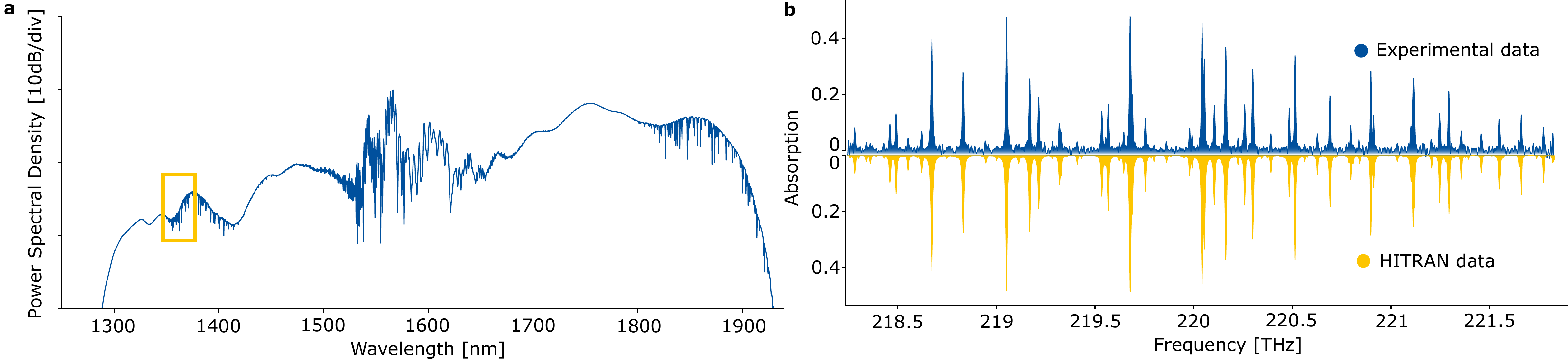}
    \caption{\textbf{Broadband spectroscopy. a} Broadband spectrum generated in normal-dispersion highly-nonlinear fiber used to perform water vapor spectroscopy around 1375 nm (yellow box). The spectrum is recorded by a grating-based optical spectrum analyzer with resolution bandwidth of 0.2 nm. \textbf{b} Experimental reconstruction of water vapor absorption features over 4 THz (blue). Comparison with the HITRAN database (yellow) shows excellent agreement.}
    \label{broad_spectro}
\end{figure*}

All phase-locked loops are digital and implemented via inexpensive, compact field-programmable gate arrays (FPGAs). The boards are programmed through the open-source python package \textit{pyrpl} \cite{neuhaus_pyrpl_2017}, replacing traditional mixers, phase and frequency counters, function generators, filters and locking modules used in analog implementations. Through a minor hard- and software modification, an external 10~MHz RF reference can be directly fed to the FPGA, providing intrinsic low-noise synchronization (the FPGA clock is generated via an internal phase-locked loop). Besides its low complexity, the digital locking scheme readily enables completely autonomous startup of the system and dynamic modification of the combs' parameters. The residual phase-noise of the beatnote between one comb and the continuous wave laser as well as the residual phase noise of one of the carrier-envelope offset frequencies are shown in Fig.~\ref{lock_scheme}b. Their respective integrated residual phase noise are $\sim 70$~mrad and $\sim 620$~mrad, integrated from 10~MHz to 100~Hz. Notably, this compact digital approach achieves a lock quality comparable to stabilization with analog electronics in a similar laser \cite{lesko_fully_2020}.

\textbf{Mutual coherence.} 
The combs' high mutual coherence manifests itself in narrow radio-frequency multiheterodyne beatnotes as shown in Fig.~\ref{lock_scheme}c,d. In this example, the beatnotes are spaced by $\Delta f_\mathrm{r} \sim 4$~Hz and their apparent 3 dB-linewidth of 1~Hz is only limited by the resolution bandwidth of the recording electrical spectrum analyzer. This sub-Hz mutual coherence permits a $>10^9$ compression factor from optical to radio-frequency domain, providing the basis for a one-to-one mapping from optical to radio-frequency domain over hundreds of THz. This spectral coverage is then only limited by the multiheterodyne detector and the spectral bandwidth of the combs. On the other hand, achieving a high compression factor enables the use of low-bandwidth heterodyne detectors, as recently demonstrated in dual-comb hyperspectral imaging \cite{voumard_ai-enabled_2020, martin-mateos_direct_2020} and photoacoustic dual-comb spectroscopy \cite{wildi_photo-acoustic_2020, friedlein_dual-comb_2020}.

\textbf{Broadband spectroscopy.} As described in the introduction, one advantage of a high-repetition DCS system is the ability to unambiguously map an optical spectrum to a RF multi-heterodyne spectrum while maintaining rapid acquisition speed. To demonstrate these characteristics, in a first experiment, a 70~THz wide spectrum (1.3 to 1.9~$\mu$m) is generated from each frequency comb in two normal-dispersion HNLFs (Fig.~\ref{broad_spectro}a). Here, $\Delta f_\mathrm{r} \approx 3$~kHz is set, which would in principle permit to cover an even wider span ($> 160$~THz), whereas in a 100~MHz system the maximal spectral span would only be $\sim$ 1.6~THz. Both broadband comb spectra traverse $\sim$10~cm of laboratory air where they experience absorption by water vapor before being combined on a 2~GHz InGaAs photodiode. As the used InGaAs-photodetector is not sensitive at wavelengths above $\sim$1800~nm where pronounced water vapor absorption features are present, we recorded the weaker absorption features at $\sim$1375~nm to validate the dual-comb signal. Figure~\ref{broad_spectro}b shows a 4~THz wide section of the measured absorption profile based on the direct multiheterodyne signal. The high-mutual coherence permits recording the dual-comb multiheterodyne spectrum directly via an RF-spectrum analyzer (in $\sim$10~s), bypassing any need for large storage and complex post-processing of the data; only a generic (high-pass) baseline correction is applied. Further, the retrieved absorption spectrum is in excellent agreement with the HITRAN database \cite{gordon_hitran2016_2017, kochanov_hitran_2016}.

\begin{figure*}
    \centering
    \includegraphics[width=0.95\textwidth]{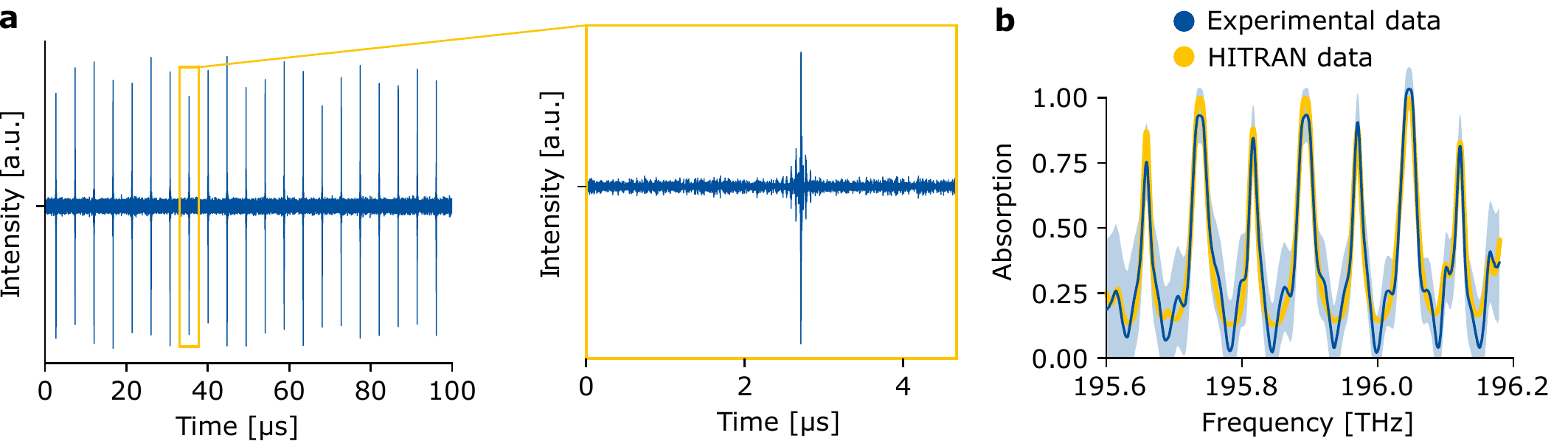}
    \caption{\textbf{Fast spectroscopy. a} Recorded series of interferograms generated with a $\sim 215$ kHz repetition rate difference and zoom-in on a single, $\sim 4.6$ $\mu$s total duration single interferogram. \textbf{b} Acetylene absorption features computed from HITRAN (yellow) and single-interferogram, experimentally retrieved absorption features (blue). The light-blue band depicts the single-interferogram standard deviation symmetrically around the single-interferogram average.}
    \label{fast_spectroscopy}
\end{figure*}

\textbf{Fast spectroscopy.}
In a second experiment, we perform spectroscopy of acetylene gas with $\mu$s-scale acquisition time. Here, the repetition rate difference is set to $\Delta f_\mathrm{r} = 215$~kHz, corresponding to a minimal, 'single-interferogram' acquisition time of $\sim$5~$\mu$s. From the Nyquist criterion, a maximal optical bandwidth of $\sim$5~THz can be covered without ambiguity, while a 100~MHz-based spectrometer would only accommodate a $\sim$50~GHz span. The dual-comb signal is optically band-pass filtered to cover a 600~GHz spectral span containing absorption features of acetylene, ensuring multi-heterodyning to the RF domain well within the Nyquist constraint. The dual-comb light is sent through a 10~cm long acetylene absorption cell at atmospheric pressure and collected on a 2~GHz InGaAs photodiode. The temporal series of the multiheterodyne interferograms recorded with an oscilloscope can be seen in Fig.~\ref{fast_spectroscopy}a along with a magnified single interferogram. A reference trace is obtained in the same experiment without acetylene in the cell. Using cross-correlation to locate the interferogram bursts, the interferogram series is interpolated on a modified time grid to ensure coherent sampling. This can alternatively be achieved on the hardware side by directly triggering the acquisition with the interferograms peaks. No other form of (phase) correction is applied.
Based on $\sim 4300$-interferograms for the absorption and reference signal an averaged absorption spectrum can be computed and is shown in Fig.~\ref{fast_spectroscopy}b. Comparison with the HITRAN database shows very good agreement. Importantly, the standard deviation of the absorption derived from only single-interferograms ($<5\mu$s acquisition time) is shown around the averaged signal.  The standard deviation is small compared to the absorption features and indicates that already a single interferogram suffices to clearly identify acetylene based on its characteristic absorption features.
Finally, we point out that operating with short acquisition times improves noise performances in dual-comb spectroscopy as low-frequency noise below $\Delta f_\mathrm{r}/2$ does not affect single interferograms. In particular the negative impact of the usual oscillator noise in the $\le 100$ kHz frequency band can be elegantly reduced when a high $\Delta f_\mathrm{r} \ge 200$ kHz is used. Moreover, a large $\Delta f_r/2$ will also improve the efficacy of phase noise correction techniques\cite{burghoff_computational_2016, ideguchi_adaptive_2014, roy_continuous_2012,  zhu_digital_2018}, which can lower noise with frequencies up to $\Delta f_r/2$.

\section{Conclusion}
A dual-frequency comb spectrometer for fast and broadband measurements based on mature 1~GHz mode-locked lasers is demonstrated. All feedback loops are implemented digitally, resulting in a compact and easy to operate setup that can readily be reconfigured. In comparison to a dual-comb spectrometer based on 100~MHz lasers, this represents a 100-fold improvement in the Nyquist criterion for the maximal spectral coverage and the minimal acquisition time. Two experiments are performed to demonstrate dual-comb characteristics that would not be possible with lower-repetition rate ($\sim$100~MHz) lasers: first, the potential for achieving a broadband ambiguity-free spectral window is demonstrated by performing spectroscopy of atmospheric water absorption features near 1375~nm over 4~THz with a 3~kHz repetition rate detuning. Second, fast spectroscopy of acetylene is performed over a 600~GHz wide optical span with an acquisition time of 5~$\mu$s, pointing towards new opportunities for observing fast transient processes such as high-throughput screening, flow-cytometry or plasma physics. In addition, high-mutual coherence with sub-Hertz multiheterodyne beatnotes is demonstrated, which in conjunction with the high-repetition rate generates new opportunities for broadband spectroscopy with low-bandwidth detectors (e.g. imaging, photo-acoustic or mid-infrared detectors). Similar to dual-comb spectroscopy, the high-repetition rate of 1~GHz is also of interest for infrared-field sampling techniques \cite{kowligy_infrared_2019} where it can speed up the measurement and relax the requirement on the lasers' noise.

\section*{Funding}
This project has received funding from the European Research Council (ERC) under the European Union’s Horizon 2020 research and innovation programme (grant agreement No 853564) from the Swiss National Science Foundation (00020\_182598) and through the Helmholtz Young Investigators Group VH-NG-1404.

\section*{Disclosures}
All authors declare no conflict of interest.

\bibliographystyle{unsrt}
\bibliography{references}

\end{document}